\documentstyle[11pt,newpasp,twoside]{article}
\def\edcomment#1{\iffalse\marginpar{\raggedright\sl#1\/}\else\relax\fi}
\reversemarginpar

\begin{document}
\title{Evolution of Dusty Disks in Nearby Young Stellar Groups}
\author{Ray Jayawardhana}
\affil{Department of Astronomy, University of California, Berkeley, CA 94720}

\begin{abstract}
Given their proximity and age differences, nearby groups of young stars
are valuable laboratories for investigations of disk evolution and diversity.
The estimated 10-Myr age of groups like the TW Hydrae Association provides a 
strong constraint on disk evolution timescales and fills a significant gap in 
the age sequence between 1-Myr-old T Tauri stars in molecular clouds and 
50-Myr-old nearby open clusters. I review the results of recent and on-going 
studies of dusty disks in three nearby groups --TW Hya, $\eta$ Cha and MBM 
12-- that suggest rapid evolution of inner disks. However, it is unlikely 
that there is a universal evolutionary timescale for protoplanetary disks, 
especially when the influence of companion stars is taken into account.
\end{abstract}

\section{Introduction}
Planetary systems are thought to form out of circumstellar disks that are 
the remnants of star formation (Shu, Adams, \& Lizano 1987). Observations of 
young pre-main-sequence (PMS) stars show that many of them are surrounded by 
optically-thick disks of solar system dimensions with masses comparable to 
or greater than the ``minimum-mass solar nebula'' of 0.01 $M_{\odot}$ (see 
Beckwith 1999 for a review). Infrared emission in excess of stellar 
photospheric fluxes provides the most readily measurable signature of such 
disks. Excesses at $\lambda \leq$ 10 $\mu$m are found in $\sim$50--90\% of the 
low-mass stars in star-forming regions (Strom et al. 1993; Lada et al. 2000). 

It has been suggested that circumstellar disks evolve from optically 
thick to optically thin structures in about 10 Myr (Strom et al. 1993). That 
transition may mark the assembly of grains into planetesimals, or clearing 
of the disk by planets. Indeed, low-mass debris disks have now been imaged 
around several main-sequence stars with ages ranging from 10 Myr to 500 Myr 
(e.g., Jayawardhana et al. 1998; Holland et al. 1998; Greaves et al. 1998). 
Since age estimates for early-type isolated main 
sequence stars are highly uncertain, however, the timescale for disk evolution 
and planet formation is poorly constrained, and may depend critically on 
the presence or absence of a close binary companion.

The nearby young stellar groups provide a unique opportunity to investigate
the evolution of circumstellar disks into planetary systems. Because of their
proximity, these groups are ideally suited for sensitive disk searches at 
near- and mid-infrared wavelengths. Furthermore, their age range of 1-50 Myrs
provides a strong constraint on disk evolution timescales and fills a 
significant gap in the age sequence between $\sim$1-Myr-old T Tauri stars 
in molecular clouds like Taurus-Auriga and Chamaeleon and the 
$\sim$50-Myr-old open clusters such as IC 2602 and IC 2391 (Jayawardhana 
2000). 

\section{TW Hydrae Association}
The TW Hydrae Association (TWA) consists of $\sim$22 co-moving stellar systems 
(Webb 2001 and references therein) with estimated 
ages of $\sim$10 Myr at a distance of 47--67 pc and dispersed over some 20 
degrees on the sky. The members are mostly late-type stars, typically K and
M spectral types, and include several binary systems as well as one 
remarkable quadruple system (HD 98800). There is only one early-type star 
(HR 4796A).

Over the past three years, we have obtained mid-infrared observations of 
TW Hya stars using the OSCIR instrument on the Keck II and CTIO 4-meter 
telescopes. We found that many of the TWA stars have little or no disk 
emission at 10 $\mu$m. Even among the five stellar systems with 10$\mu$m 
excesses, most show some evidence of inner disk evolution. We imaged a 
spatially-resolved dust disk around the young A star HR 4796A (Jayawardhana 
et al. 1998; Telesco et al. 2000; Fig. 1). The surface brightness 
distribution of the disk is consistent with the presence of an inner disk 
hole of $\sim$50 AU radius, as was first suggested by Jura et al. (1993) 
based on the infrared spectrum. The SEDs of HD 98800 and Hen 3-600A also 
suggest possible inner disk holes (Jayawardhana et al. 1999a; 1999b). The 
excess we detected for CD -33$^{\circ}$7795 is modest, and could well be due 
to a faint companion. Only TW Hya appears to harbor an optically-thick, 
actively accreting disk of the kind observed in $\sim$1-Myr-old classical
T Tauri stars; it is the only one with a large H$\alpha$ equivalent
width (-220 \AA). Recent polarimetric images using the adaptive optics
system on Gemini North confirm a nearly face-on disk around TW Hya (D. 
Potter, private communication), previously detected in scattered light 
by WFPC2 (Krist et al. 2000) and NICMOS (Weinberger et al. 1999) on HST and 
in millimeter emission by the VLA (Wilner 2001). However, even in
TW Hya, there is evidence for dust settling in the inner disk (D'Alessio
2001) and lower accretion rates (Muzerolle et al. 2001) compared to classical 
T Tauri stars (CTTS) in Taurus.

\begin{figure}
\plotfiddle{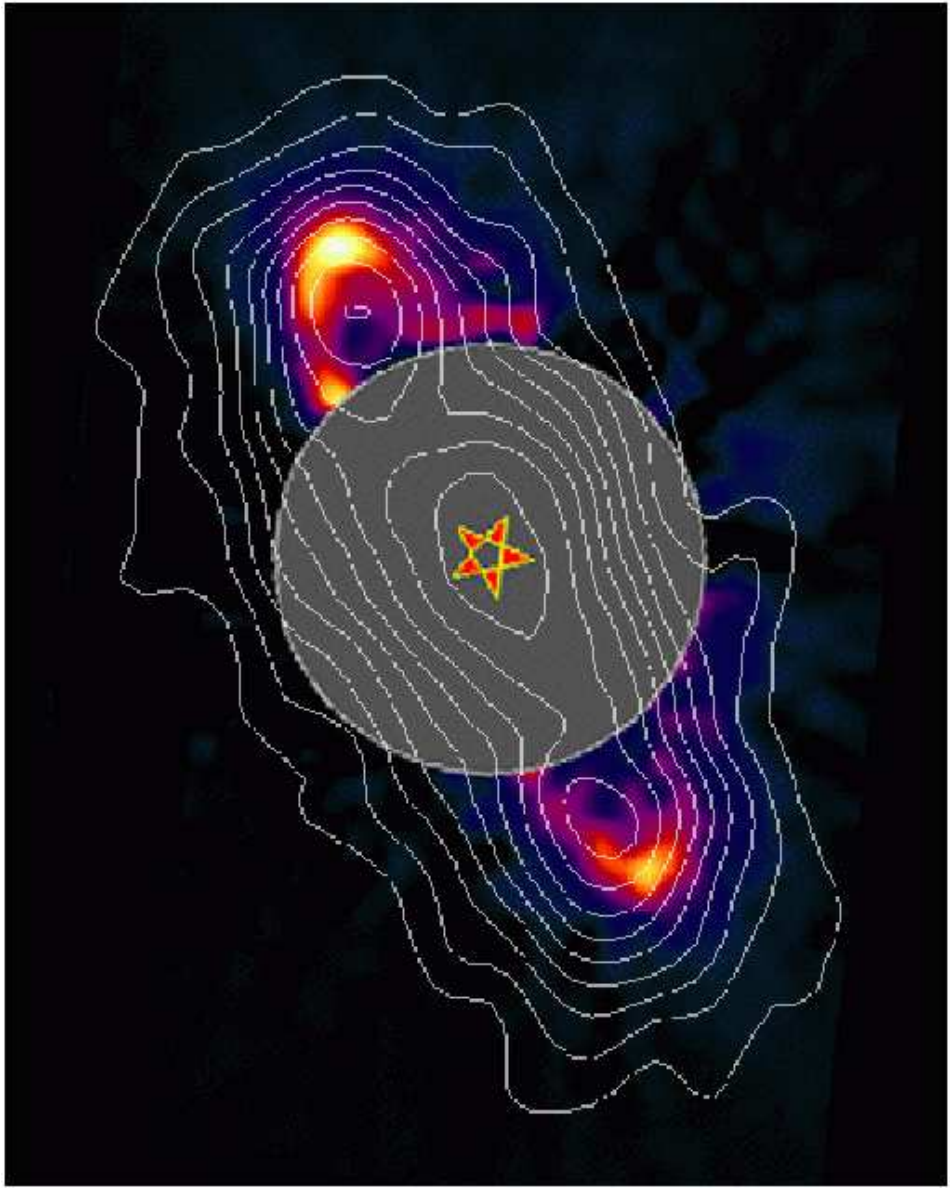}{4.8in}{0}{70.0}{70.0}{-210.0}{-180.0}
\vspace{2.2in}
\caption{Overlay of Keck/OSCIR 18.2$\mu$m contours on the 1.1$\mu$m HST/NICMOS 
coronagraphic image of HR 4796A disk. From Telesco et al. (2000).}
\end{figure}

If most TWA stars are $\leq$10 Myr old, the above results suggest that 
their inner disks have already depleted either through coagulation of dust 
or accretion on to the central star.  The fact that only one (TW Hya) out of 
16 systems we observed shows classical T Tauri characteristics (compared to
$\sim$50--90\% of $\sim$1-Myr-old stars in star-forming regions) argues for
rapid evolution of inner disks in pre-main-sequence stars. However, it is 
unlikely that there is a universal evolutionary timescale for protoplanetary 
disks, especially when the influence of companion stars is taken into account.
For example, we have detected thermal emission from a dusty disk around the 
primary, but not the secondary, in the Hen 3-600 binary system (Jayawardhana
et al. 1999a). Comparison with the median spectral energy distribution of 
classical T Tauri stars suggests that the disk around Hen 3-600A may be 
truncated by the secondary. Sitko et al. (2000) find that the 10$\mu$m 
silicate emission feature in TW Hya is similar to that of other CTTS whereas 
it is weaker in HD 988000 and almost non-detectable in HR 4796A.

\section{$\eta$ Chamaeleontis}
$\eta$ Cha is a compact cluster of about a dozen stars at $\sim$97 pc,
far from any obvious molecular material (Mamajek, Lawson \& Feigelson 1999).
Among the 12 previously known members of the $\eta$ Cha cluster, only
2 have H$\alpha$ equivalent widths larger than 10 \AA~, suggestive of disk 
accretion. Preliminary results of our February 2001 mid-infrared observations 
using the TIMMI2 instrument on the ESO 3.6-meter telescope confirm that 
none of these 12 stars have large excesses consistent with optically-thick
disks. However, Lawson (2001) reports the identification of a new CTTS 
assciated with the cluster, and another member (RECX 11) is listed in
the IRAS Faint Source Catalog as showing a far-infrared excess. Other $\eta$ 
Cha stars may have optically thin disks with central cavities, as is the case 
in several TWA stars, but our analysis is not yet complete.

\section{MBM 12}
Based on {\it ROSAT} detections and ground-based follow-up optical 
spectroscopy, Hearty et al. (2000) identified 8 late-type young stars 
in the high-latitude cloud MBM12 as well as two main-sequence stars in the 
same line-of-sight which may or may not be related. While there is some
uncertainty in the distance to MBM 12 (Luhman 2001), it is likely to be a 
younger group than TW Hya and $\eta$ Cha, perhaps only 1-3 Myr in age.

Recently, we have conducted the first investigation of protoplanetary disks 
in the MBM 12 group, using mid-infrared imaging with Keck and UKIRT and 
optical spectroscopy with the Kitt Peak 4-meter telescope (Jayawardhana
et al. 2001). The ($K-L$) and 
($K-N$) colors we derived unambiguosly show significant mid-infrared excess 
from six MBM 12 stars  --LkH$\alpha$ 262, LkH$\alpha$ 263,
LkH$\alpha$ 264, E02553+2018, RXJ0258.3+1947 and S18. In all six cases,
the colors are consistent with thermal emission from optically-thick 
inner disks. The other two PMS stars --RXJ0255.4+2005 and RXJ0306.5+1921--
do not show a measurable mid-infrared excess within the photometric errors,
allowing us to rule out such disks. HD 17332 and RXJ0255.3+1915, the two
main-sequence stars in the line-of-sight to MBM 12, also lack evidence of
warm circumstellar material. Four of the objects --LkH$\alpha$ 262, 
LkH$\alpha$ 263, LkH$\alpha$ 264 and E02553+2018-- show two components in 
the H$\alpha$ line, and are probably close binaries (Fig. 2). An H$\alpha$ 
equivalent width greater than 10 \AA~ is generally considered to 
be the accretion signature of a classical T Tauri star with a circumstellar 
disk (Herbig \& Bell 1988). A smaller equivalent width signifies chromospheric 
activity, but no accretion. Therefore, mid-infrared excess we measure should 
be correlated with large H$\alpha$ line widths. Indeed, this generally holds 
true for MBM 12 stars with one notable exception: E02553+2018 has a large 
mid-infrared excess but weak H$\alpha$ line emission. We tentatively suggest 
that E02553+2018 is a candidate for harboring circumbinary dust. 

\begin{figure}
\plotfiddle{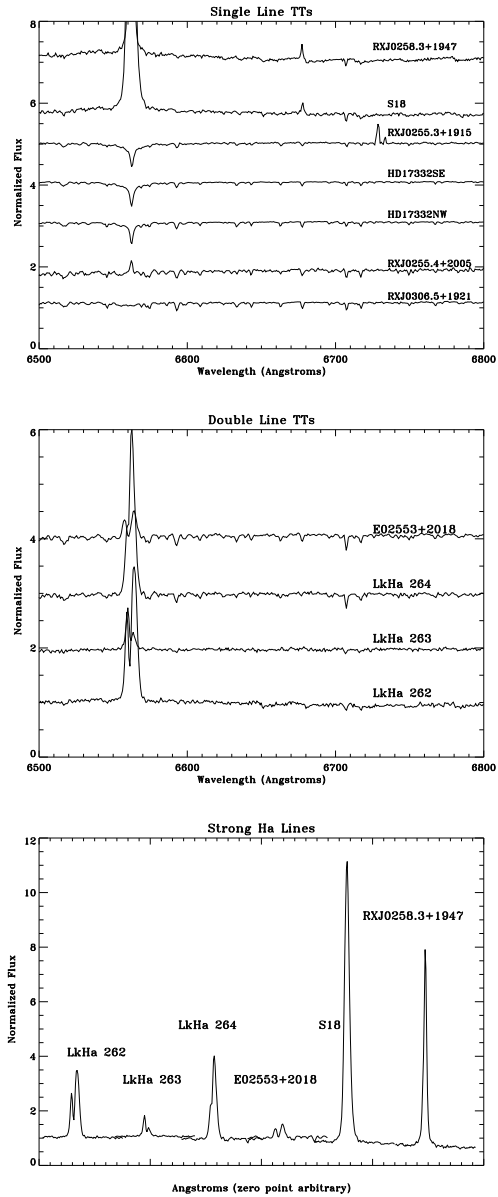}{4.8in}{0}{70.0}{70.0}{-220.0}{-110.0}
\vspace{0.8in}
\caption{The optical spectra of single-line ({\it top panel}) and
double-line ({\it center panel}) systems among MBM 12 stars. Bottom 
panel shows detail of the H$\alpha$ lines of strong emitters. From 
Jayawardhana et al. (2001).}
\end{figure}

The disk fraction we find in MBM 12 --75\%-- falls in the middle of the 
range reported for other star-forming regions like Taurus and Trapezium, but
is significantly higher than in TWA and $\eta$ Cha. Preliminary results of 
recent 3mm OVRO observations of MBM 12 stars suggest that their disk masses 
fall in the low-mass tail of the Taurus CTTS distribution (Hogerheijde et al. 
2001).

\acknowledgments
RJ holds a Miller Research Fellowship. This work was supported in part by
a NASA grant administered by the AAS.

\end{document}